# Observation of non-superconducting phase changes in LuH$_{2\pm x}$N$_y$

Xiangzhuo Xing[1,2,#], Chao Wang[1,2,#], Linchao Yu[1], Jie Xu[1], Chutong Zhang[1], Mengge Zhang[1], Song Huang[1], Xiaoran Zhang[1], Bingchao Yang[1,2], Xin Chen[1,2], Yongsheng Zhang[1,2], Jian-gang Guo[3], Zhixiang Shi[4], Yanming Ma[5,6,7], Changfeng Chen[8] and Xiaobing Liu[1,2,*]

[1] *Laboratory of High Pressure Physics and Material Science (HPPMS), School of Physics and Physical Engineering, Qufu Normal University, Qufu 273165, China*

[2] *Advanced Research Institute of Multidisciplinary Sciences, Qufu Normal University, Qufu 273165, China*

[3] *Beijing National Laboratory for Condensed Matter Physics, Institute of Physics, Chinese Academy of Sciences, Beijing 100190, China*

[4] *School of Physics, Southeast University, Nanjing 211189, China*

[5] *Innovation Center for Computational Methods & Software, College of Physics, Jilin University, Changchun 130012, China*

[6] *State Key Laboratory of Superhard Materials, Jilin University, Changchun 130012, China*

[7] *International Center of Future Science, Jilin University, Changchun 130012, China*

[8] *Department of Physics and Astronomy, University of Nevada, Las Vegas, Nevada 89154, USA*

[#] *These authors contributed equally to this work.*

**The recent report of near-ambient superconductivity in nitrogen doped lutetium hydride[1] has triggered a worldwide fanaticism and raised major questions about the latest claims. An intriguing phenomenon of color changes in pressurized samples from blue to pink to red was observed and correlated with the claimed superconducting transition, but the origin and underlying physics of these color changes have yet to be elucidated[2-7]. Here we report synthesis and characterization of high-purity nitrogen doped lutetium hydride $LuH_{2\pm x}N_y$ with the same structure and composition as in the main phase of near-ambient superconductor[1]. We find a new purple phase of $LuH_{2\pm x}N_y$ between blue and pink phase, and reveal that the sample color changes likely stem from pressure-driven redistribution of nitrogen and its interaction with the $LuH_2$ framework. No superconducting transition is found in all blue, purple, pink and red phases at temperatures 1.8-300 K and pressures 0-30 GPa. Instead, we identify a notable temperature-induced resistance anomaly of structural and/or electronic origin in $LuH_{2\pm x}N_y$, which is most pronounced in the pink phase and may have been erroneously interpreted as a sign of superconducting transition. This work establishes key benchmarks for nitrogen doped lutetium hydrides, allowing an in-depth understanding of the novel pressure-induced phase changes.**

A recent study reported near-ambient superconductivity in a nitrogen doped lutetium hydride[1]. High-temperature superconductivity has been predicted even realized in metallic hydrogen and hydrogen-rich compounds (such as sulfur hydride[8], rare-earth hydrides[9-11] and alkaline-earth hydrides[12,13]), but megabar pressures are required to stabilize structures. The superconductivity in nitrogen doped lutetium hydride was achieved at much reduced pressures of 0.3-3 GPa, with maximum critical temperature $T_c$=294 K at 1.3 GPa. This discovery has sparked tremendous interest in the scientific community and beyond; ensuing studies quickly followed[2-4,14-18], which have thus far found no evidence supporting superconductivity in Lu-N-H systems.

It was reported that the presence of near-ambient superconductivity in the nitrogen doped lutetium hydride coincided with a visual color change of the sample from blue (0-0.3 GPa) through pink (0.3-3 GPa) to red (>3 GPa)[1]. Conspicuously, the superconducting state was only found in the pink phase. This is an intriguing phenomenon since all previously reported high-temperature superconductors show dark or black color in hydrogen-rich metallic systems[9-13]. Meanwhile, similar color changes were observed in $LuH_2$ at pressures ranging from 2.5 GPa and 5 GPa[2], but absent in lutetium hydride samples doped with nitrogen at pressures of 0-6 GPa[3]. No superconducting transition was detected in $LuH_2$ sample[2] or the blue phase of the $LuH_{2\pm x}N_y$ sample[3] at temperatures from 300 K to 10 K. Since all the samples in three reported works share the same crystal structure, their main distinctions are the sample color and nitrogen contents. Producing and probing high-quality nitrogen doped lutetium hydrides is crucial to elucidating pertinent phenomena and the underlying mechanisms.

In this work, we synthesized pure bulk samples of nitrogen doped lutetium hydrides by high pressure and high temperature (HPHT) method. The samples exhibit uniform shinning blue color and have the same crystal structure and well-distributed nitrogen content as in previously reported samples[1]. Our *in-situ* high-pressure experiments revealed reversible color changes of nitrogen doped lutetium hydrides from shining blue to dark blue to purple and pink, finally into red. The critical pressures for these color changes are sensitive to the used pressure media, but the overall trends are consistent among all the cases. Our electrical measurements did not detect any signals of a superconducting transition at temperatures from 300 K to 1.8 K under pressures up to 30 GPa. There is an abnormal feature above 200 K during the warming-up electrical measurements on the pink and red phases, but the sample retains normal metallic behavior during the cooling-down measurements. The raw data taken in the warming-up measurements of the reported near-ambient superconductors[1] exhibit the same anomalous resistance, which was interpreted as a sign of superconductivity after a background subtraction was applied. Our results from both warming-up and cooling-down cycles on the same sample show unambiguous evidence that there is no correlation between the pink phase and the claimed near-ambient superconductivity in nitrogen doped lutetium hydrides since otherwise the data obtained during the cooling-down cycle also should capture the same basic physics associated with the superconducting transition. Also, the fact the same resistance anomaly is also seen in the red phase, which was recognized to be non-superconducting[1], further reinforces our conclusion about the lack of a superconducting transition in the sample.

**Synthesis, structure and composition of the produced $LuH_{2\pm x}N_y$ sample**

We performed HPHT experiments for synthesis of nitrogen doped lutetium hydrides (see Methods for details and a schematic diagram in Extended Data Figure 1). Figure 1a shows the X-ray diffraction (XRD) pattern on a polished smooth surface of the shining blue sample (top inset) by a diffractometer with wavelength of 1.5406 Å. All the dominant peaks can be well indexed as a cubic structure with $Fm\bar{3}m$ space group, and no impurity phases such as $LuN_{1-\delta}H_\epsilon$ and $Lu_2O_3$ phases are detected except for a small amount of lutetium left during the surface polishing in preparation for XRD measurements. The crystal structure is consistent with that of $LuH_2$[19] and the reported near-ambient superconductors[1]. The sharp diffraction peaks indicate a better crystalline quality in our HPHT products compared to previous samples[1].

High-resolution transmission electron microscopy (HRTEM) measurements (Fig. 1b) reveal that the Lu atoms arrange in the (200), (111) and ($1\bar{1}\bar{1}$) orientations with the lattice spacing of 0.252 nm, 0.288 nm and 0.299 nm, respectively. The selected area electron diffraction (SAED) pattern along the [$\bar{1}$10] zone-axis is shown in Fig. 1c, which is consistent with the result of our XRD measurement. The lattice parameter $a$ is determined to be 5.040 Å, which is much smaller than that of LuH$_3$[2], but is comparable to that of recently reported LuH$_{2\pm x}$N$_y$ samples[3]. Thus, we attribute the composition of our HPHT produced samples as LuH$_{2\pm x}$N$_y$.

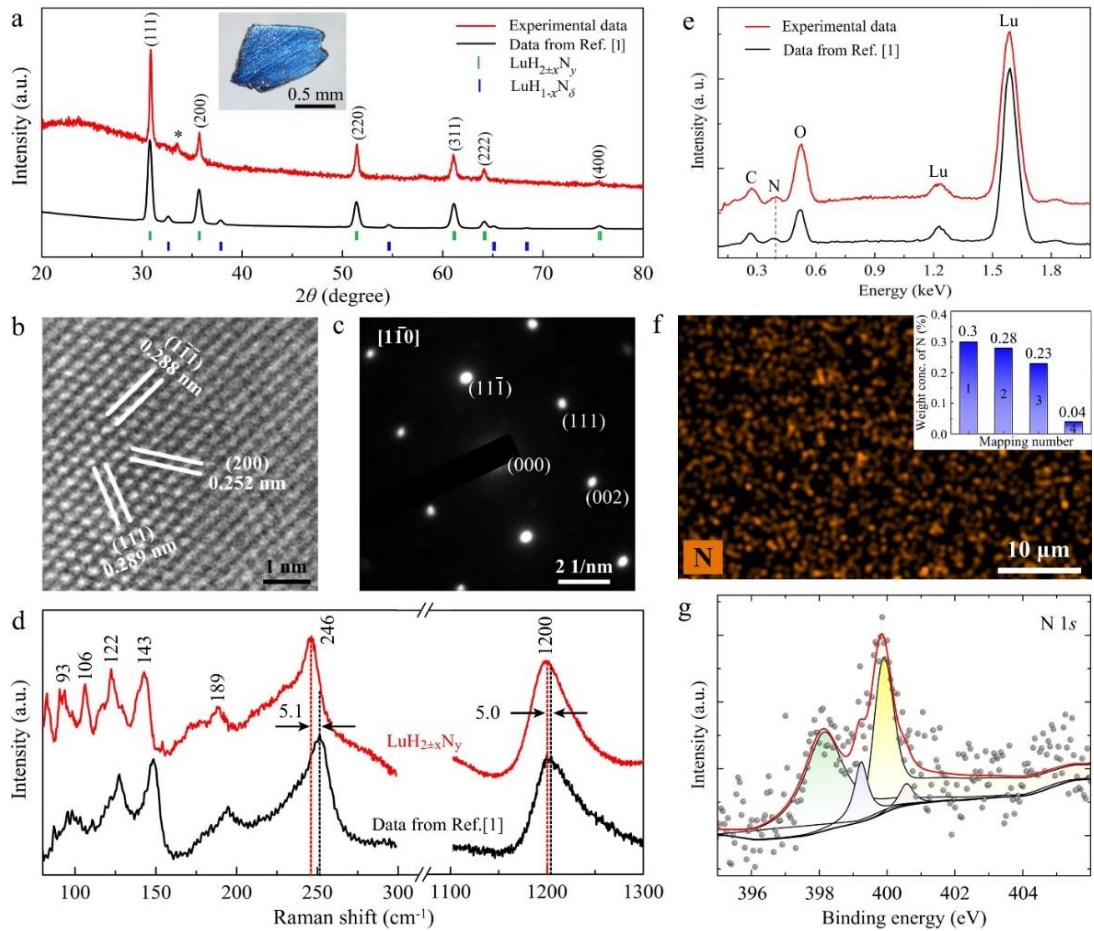

**Fig. 1 | Structure and composition of the produced nitrogen doped lutetium hydride sample.** The reported data of the near-ambient superconducting sample is plotted as black lines (bottom) for comparison. **a**, Typical XRD pattern (top) of our produced nitrogen doped lutetium hydride sample. The small peak marked by the star is from the residual lutetium. Top inset image is the measured bulk sample, around 1 mm in size. **b** and **c**, HRTEM image of the LuH$_{2\pm x}$N$_y$ sample and the SAED pattern along the [$\bar{1}$10] zone-axis. **d**, Typical Raman spectra of LuH$_{2\pm x}$N$_y$ under ambient conditions. **e**, A representative EDS spectrum of the produced LuH$_{2\pm x}$N$_y$ sample. The carbon and oxygen peaks are from the tape used for holding a tiny sample for EDS measurements. **f**, Mapping image for nitrogen elements. Top inset is the average weight concentration of nitrogen from the

EDS mapping measurements at four different areas, 50×35 μm in size. **g**, XPS spectra of the N 1*s* core level.

Figure 1d displays typical Raman spectrum of the $LuH_{2\pm x}N_y$ samples with a 532 nm laser excitation. The obtained spectrum is practically identical to that of the previously reported sample[1] with characteristic peaks at 83 cm$^{-1}$, 93 cm$^{-1}$, 106 cm$^{-1}$, 122 cm$^{-1}$, 143 cm$^{-1}$, 246 cm$^{-1}$ and 1200 cm$^{-1}$. There is a nearly uniform down-shift of about 5 cm$^{-1}$ for all the Raman peaks in our samples, which is likely due to a lack of $LuH_{1-x}N_\delta$ impurities seen in previous samples[1]. The peaks at 246 cm$^{-1}$ and 1200 cm$^{-1}$ are close to those in $LuH_2$[2], thus likely coming from Lu-H, while the peaks at the lower wavenumbers below 150 cm$^{-1}$ are only observed in nitrogen doped lutetium hydrides[1] that can be assigned to N related vibrational modes. The scanning electron microscope and mapping measurements by Raman spectroscopy show a uniform polycrystalline structure of the produced $LuH_{2\pm x}N_y$ samples (Extended Data Fig. 2 and Fig. 3).

Energy dispersive X-ray spectroscopy (EDS) spectrum in Fig. 1e shows clear evidence for the incorporation of nitrogen in our $LuH_{2\pm x}N_y$ samples, while the EDS mapping results indicate macroscopically uniform nitrogen distribution (Fig. 1f). We calculated the averaged nitrogen content in our $LuH_{2\pm x}N_y$ samples based on four EDS mapping results, and three are around 0.2-0.3 wt.% and one is very low about 0.04 wt.%. We also randomly measured 10 spots in the sample, obtaining a maximum value of 0.75 wt.%, at the same level with previously reported nitrogen doped lutetium hydrides[1,3], indicating that these samples have similar nitrogen contents. X-ray photoelectron spectroscopy (XPS) measurements further confirm incorporation of nitrogen in the samples (Fig. 1g and Extended Data Fig. 4).

**Pressure-induced color changes in the $LuH_{2\pm x}N_y$ sample**

To investigate the color evolution of the nitrogen doped lutetium hydrides, we loaded the produced $LuH_{2\pm x}N_y$ samples into standard diamond anvil cells without any medium or with different pressure media of nitrogen gas, water, silicon oil, solid NaCl in the sample chambers. Our experimental results demonstrate that the color change with increasing pressure can be clearly seen in all the studied stress environments (see Extended Data Fig. 5-9).

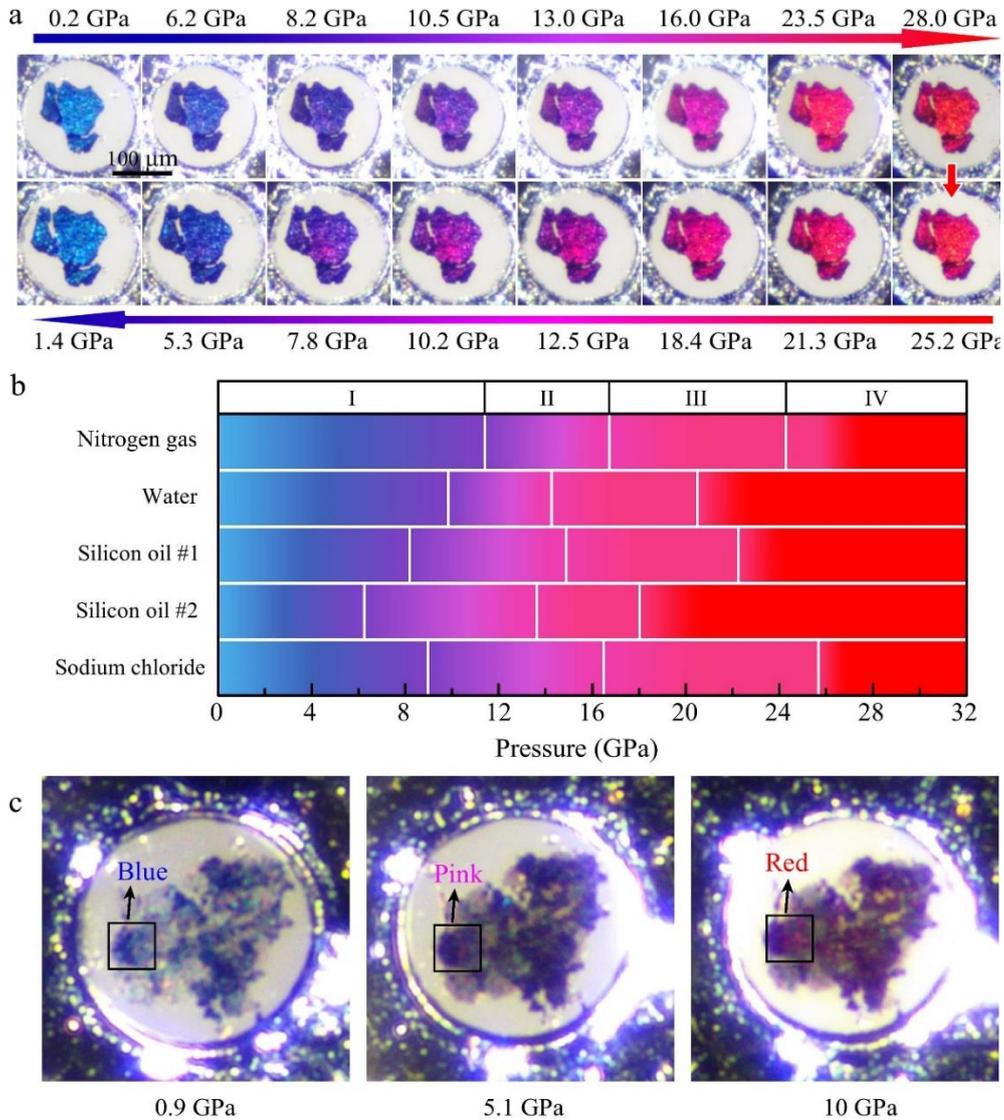

**Fig. 2 | Evolution of color changes of nitrogen doped lutetium hydrides with varying pressures**. **a**, Optical images of pressure-induced color changes of the produced $LuH_{2\pm x}N_y$ samples during compression (upper images) and decompression (down images) processes in a DAC chamber with water as pressure medium. **b**, Phase diagram of phase I (blue), phase II (purple), phase III (pink) and phase IV (red) of the $LuH_{2\pm x}N_y$ samples in different pressure media. Two runs were carried out in silicon oil medium with Re-gasket thickness of 32 and 44 μm, respectively. The sample chamber with thick gasket underwent expansion, creating shear stresses, which helped reduce the critical pressures of color changes in the $LuH_{2\pm x}N_y$ sample. **c**, Optical

images of the samples showcasing reduced critical pressures of the $LuH_{2\pm x}N_y$ samples mixed with nanodiamonds by internal stress in the NaCl pressure medium.

Figure 2a shows a systematic evolution pattern of color change of the $LuH_{2\pm x}N_y$ sample in the silicon oil medium over a wide pressure range. With pressure increasing to 6.2 GPa, the initial shining blue color turns to royal blue, then to purplish blue at 8.2 GPa, further to purple at 10.5 GPa. The intriguing pink phase starts to show up at 16 GPa and remains over a span of about 7 GPa until the vivid red color appears at 23.5 GPa. During the decompression process, the color evolution is completely revisable, gradually from vivid red back to its original shinning blue after pressure is fully released. Figure 2b depicts the critical pressure and stable region for the color changes in $LuH_{2\pm x}N_y$ samples in different pressure media. Our work reveals a previously unseen purple phase that is present in all the cases studied in our work, but was missed by Dasenbrock-Gammon et al. because of the very narrow pressure range for the transition from blue phase to pink phase in their work[1]. Our results establish a pressure-induced phase change sequence as indicated by their colors from phase I (blue) to phase II (purple), then phase III (pink), and eventually phase IV (red).

The overall trend of pressure induced color variation in our work is fully consistent with that of the reported near-ambient superconductors[1], but the critical pressures for color changes are higher in our work. This phenomenon is also observed in another independent work, which did not observe any color change of the nitrogen doped lutetium hydrides at pressures up to 6 GPa[3]. The much lower critical pressures in the previous work[1] can be attributed to the composite nature of their samples containing dominant A and minority B phases with different lattice parameters, and these mixed phase structures can introduce large and complex internal stresses between the constituent components under compression. Such complex stresses could lead to significant changes, including large reduction of critical pressure for phase transitions[20-23]. This also explains the distinct critical pressures and stable regions of the differently colored phase in different pressure media that exert different stresses on the samples. To further verify this point, we introduce stresses by adding a small amount of nanodiamonds into $LuH_{2\pm x}N_y$ powers in the NaCl pressure medium, as illustrated in Fig. 2c. Consequently, the pink and red phases occur at significantly reduced pressures of 5.1 GPa and 10.0 GPa, respectively, suggesting that the presence of minority B phase mixed within dominant A phase in the samples of the previously report work[1] is effective in pushing the critical pressure for the pink phase down to 0.3 GPa.

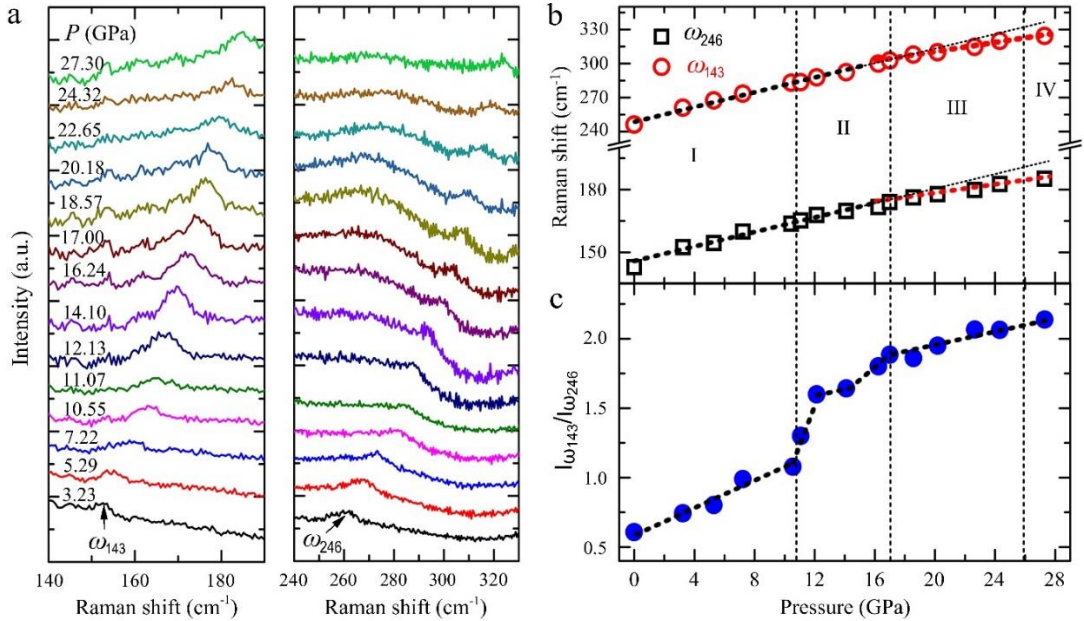

**Fig. 3 | *In-situ* Raman spectrum of nitrogen doped lutetium hydrides at changing pressures**. **a**, Pressure-dependent Raman spectrum of the produced $LuH_{2\pm x}N_y$ sample. **b**, Pressure-dependent shift of Raman peaks at 143 cm$^{-1}$ and 246 cm$^{-1}$. **c**, Intensity ratio between the two Raman peaks of 143 cm$^{-1}$ and 246 cm$^{-1}$ with increasing pressure.

## *In-situ* Raman spectrum of the $LuH_{2\pm x}N_y$ sample under compression

Figure 3a shows pressure-dependent Raman spectrum of the produced $LuH_{2\pm x}N_y$ sample up to 27.3 GPa in the nitrogen gas medium, which has gone through the whole color change sequence. The strongest 1200 cm$^{-1}$ peak (Fig. 1d) quickly overlap with the Raman peaks at 1332 cm$^{-1}$ from diamond anvils under compression, so we selected the second (N related vibration) and third (Lu-H related vibration) peak for comparison. Results in Fig. 3b show that both of these peaks gradually move upward in frequency at nearly the same rate, and the rate of decline decrease when it gets into the pink and red phases. It is clearly seen that there is a significant enhancement in the intensity of the N related peak with increasing pressure, while the Lu-H related peak loses intensity at rising pressure. An interesting observation is an accelerated rising rate for the relative intensity of the N related peak to that of the L-H related peak in the purple phase region before the appearance of the pink phase (Fig. 3c). The intensity of N the related vibration is about twice that of the Lu-H related vibration in the pink and red phase. These results offer strong evidence that the pressure conditions have major effects on the redistribution of nitrogen and interaction between nitrogen and the structural frame of lutetium hydrides. This may be a key reason for the gradual color evolution from its original blue to the ultimate red under compression.

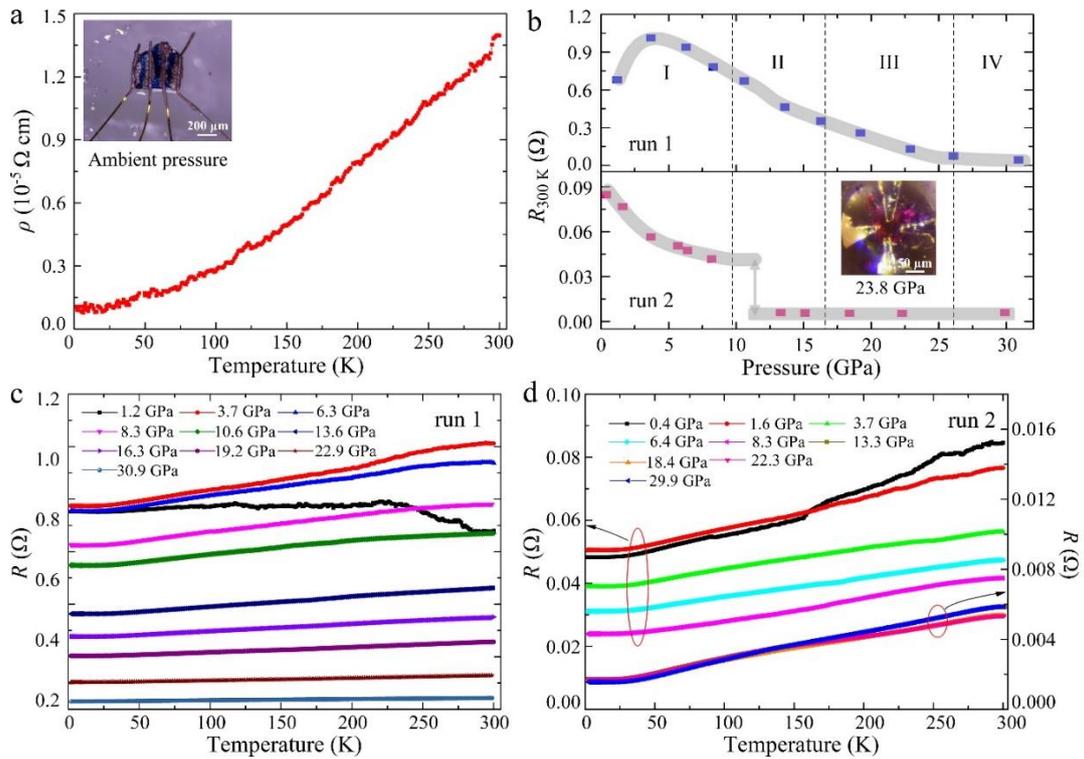

**Fig. 4 | Temperature dependent electrical resistance of the LuH$_{2\pm x}$N$_y$ at changing pressures**. **a**. Temperature dependence of resistivity at ambient pressure. Top inset shows the optical micrograph of LuH$_{2\pm x}$N$_y$ sample with gold electrodes attached by silver paste. **b**, Room temperature resistance at different pressures. **c** and **d**, Evolution of the resistance with pressure for run 1 and run 2, respectively.

# Electrical transport measurements of the produced LuH$_{2\pm x}$N$_y$ samples

We performed electrical transport measurements in the four phases over the temperature range of 1.8-300 K. Results show that LuH$_{2\pm x}$N$_y$ exhibits typical metallic behaviors at ambient pressure (Fig. 4a). We further measured the temperature dependent resistance in two runs with samples taken from our synthesized bulk LuH$_{2\pm x}$N$_y$ and using NaCl and Pt as pressure medium and electrodes, respectively. Figure 4b shows pressure evolution of resistance of LuH$_{2\pm x}$N$_y$ at room temperature. In run 1, under rising pressure, the resistance first increases until pressure reaches 3.7 GPa, then gradually decreases up to pressure of 30.9 GPa. In run 2, the resistance decreases with rising pressure up to 8.3 GPa and drops by an order of magnitude between 8.3 and 13.3 GPa, then stays nearly constant to the highest measured pressure of 29.9 GPa. Such diverse behaviors of resistance under pressure likely stems from the variation of the nitrogen contents in different samples taken from different parts of the original synthesized sample, see Fig. 1f. In Fig. 4c and 4d, we present the resistance during the cooling-down measurements, and the results show no sign of superconductivity in all four phases (blue to red) from 300 K to 1.8 K.

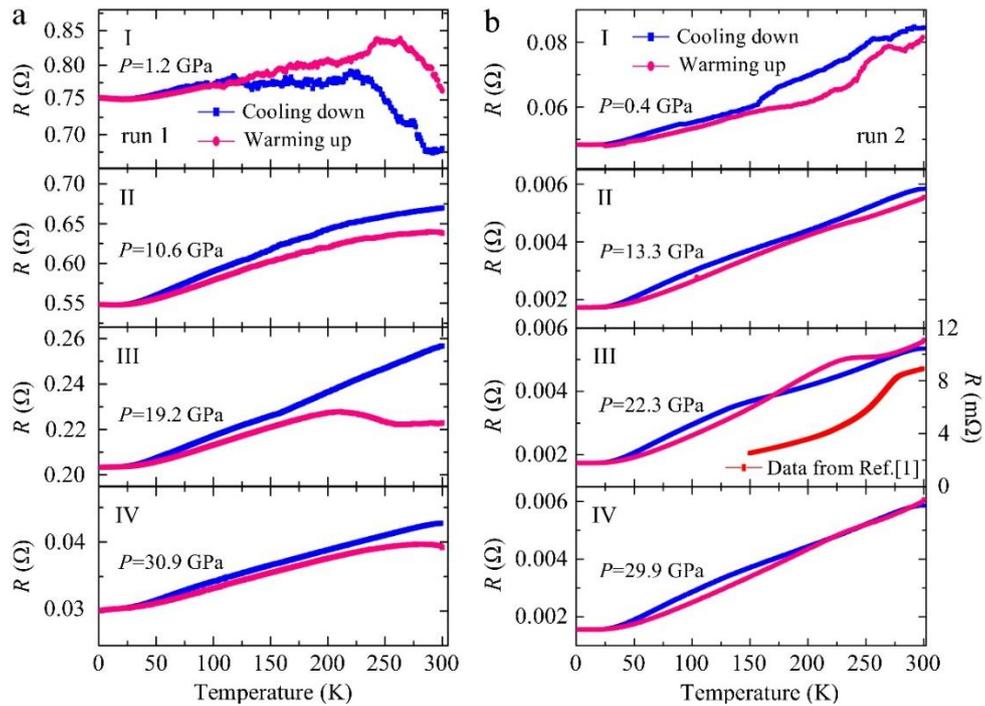

**Fig. 5 | Contrasting *R*(T) curves obtained in cooling-down and warming-up measurements.
a**, Temperature dependence of resistance in run 1 for phase I at 1.2 GPa, phase II at 10.6 GPa, phase III at 19.2 GPa and phase IV at 30.9 GPa. **b**, Temperature dependence of resistance in run 2 for phase I at 0.4 GPa, phase II at 13.3 GPa, phase III at 22.3 GPa and phase IV at 29.9 GPa. The raw resistance data on the pink phase (taken at lower pressure) from the recent work reporting on ambient-condition superconductivity[1] are included in (b, panel III) to show similar hump observed in our warming-up curve, despite a shift in temperature where the turning point of the curve is located, which may have been caused by the effect of the internal stress.

The measured resistance curve should exhibit the same trend of variation with changing temperature during cooling-down or warming-up process with same cooling/warming rate if no temperature-driven phase transition occurs. Generally, data collected during warming-up are used for analysis because of the more homogeneous thermal equilibrium than in the cooling-down measurements. Surprisingly, we observed contrasting behaviors in the measured *R*(T) curves of $LuH_{2\pm x}N_y$ during the cooling-down and warming-up electrical measurements (see Extended Data Fig. 10 and Fig. 11), and the contrast is especially pronounced in the pink phase (III) that was claimed to host near-ambient superconductivity[1]. At 1.2 GPa, the *R*(T) curve for phase I has a hump at relatively high temperatures (Fig. 5a) during both cooling-down and warming-up process, and this behavior was also seen in $LuH_2$, thus considered intrinsic to the $LuH_2$ crystal[2]. This behavior was suppressed by increasing pressure at 3.7 GPa and stays suppressed until the early part of phase II (purplish blue) at 8.3 GPa. As the $LuH_{2\pm x}N_y$ sample turns purple well inside phase II at 10.6 GPa, the hump structure recurs, but intriguingly, only during the warming-up measurements, and the hump structure becomes more pronounced through the pink phase III and persists into the red phase IV. The red phase was recognized by previous work[1] to be non-superconducting, which is confirmed by our measured resistance data. These results consistently show that the hump in the resistance curves in the pink and red phases are not signs of superconductivity in the sample. We also measured *R*(T) with an applied magnetic field of 3 T, the results show no shift in the critical pressure marking the turning point in the resistance curve, and this lack of response of resistance to applied magnetic field excludes the possibility of a superconducting transition for this abnormal behavior (see Extended Data Fig. 10j).

Data collected from run 2 also show different evolution of the resistance in cooling-down and warming-up process (Fig. 5b). Most interestingly, in the region of purple phase II (above 13.3 GPa) and pink phase III (22.3 GPa) (Extended Data Fig. 11 h-11j), a plateau in the resistance curve develops and grows with rising pressure starting at about 260 K during the warming-up measurements, followed by a relatively sharp drop with reducing temperature; however, no zero-resistance state was observed down to 1.8 K. Such a feature significantly weakens when the sample turns to red color (Phase IV). These results suggest the presence of a temperature-driven structural or electronic change in $LuH_{2\pm x}N_y$, possibly caused by the redistribution of nitrogen and its interaction with the LuH framework, as indicated by the Raman spectra shown in Fig. 3a. The onset temperature of the hump structure in $LuH_{2\pm x}N_y$ is sensitive to pressures and rises in the pink phase with increasing pressure (Extended Data Fig. 12).

## Conclusion

We have synthesized high-purity nitrogen doped lutetium hydrides ($LuH_{2\pm x}N_y$) using HPHT method, and the obtained samples exhibit the same crystal structure, composition and overall color changing trends as those of the recently reported near-ambient superconductor[1]. A new purple phase of $LuH_{2\pm x}N_y$ is discovered to exist during transformation from blue to pink phase under compression. Our extensive and systematic resistance measurements indicate that there is no superconducting transition in all the blue, purple, pink and red phases of $LuH_{2\pm x}N_y$ samples at temperatures of 1.8-300 K and pressures from 0.4 to 30 GPa. These results demonstrate that the pressure/temperature-driven redistribution of nitrogen and its interaction with the $LuH_2$ framework plays a crucial role in the remarkable visual color changes and anomalous electrical resistance behaviors seen in the experiments. The present work has shown unambiguous evidence that there is no correlation between the pink phase and near-ambient superconductivity in nitrogen doped lutetium hydrides, and further efforts should focus on elucidating the origin of the intriguing emergent N related bonding changes that may drive the sample color changes and the associated resistance anomalies.

**Acknowledgements** X.L. would like to thank the whole HPPMS team for the hard work. The authors would like to thank Prof. Xiaofeng Xu for fruitful discussion. This work was partly supported by the National Natural Science Foundation of China (Grant Nos. 12204265, 11974208 and 52172212), the Natural Science Foundation of Shandong Province (Grant Nos. ZR2020YQ05,

ZR2022QA040, 2022KJ183 and 2019KJJ020) and the Strategic Priority Research Program (B) of the Chinese Academy of Sciences (Grant No. XDB25000000).

**Author contributions** X.L. and X.X. designed the experiments and gathered data. X.L., X.X., C.C. and Y.M. analyzed the data, organized the results and wrote the manuscript. X.X., L.Y., S.H. and X.L. synthesized the samples. X.X, C.W., J.X., M.Z., C.Z., X.Z. and P.L. performed sample characterization and high-pressure measurements. B.Y., X.C., Y.Z., J.G. and Z.S. provided discussions. All authors contributed to editing and improving the manuscript.

**Author Information** The authors declare no competing interests.

**Correspondence and requests for materials** should be addressed to X.L. (xiaobing.phy@qfnu.edu.cn)

**Supplementary Information** is available for this paper.

## METHODS

**Starting materials and HPHT synthesis.** The polycrystalline samples of $LuH_{2\pm x}N_y$ were synthesized using a China-type cubic-type high pressure apparatus. The high-pressure cell used in this study has two layers (layer 1 and layer 2) that were separated by a BN thin plate, as depicted by a schematic diagram in Extended Data Fig. 1. In layer 1, the Lu pieces (99.9 wt.% purity) with silver color were placed as the precursors. Layer 2 was filled with the mixture of $NH_4Cl$ (Aladdin 99.99 wt.% purity) and $CaH_2$ (Aladdin 98.5 wt.% purity) in a molar ratio of 2:8 that was used as the source of nitrogen and hydrogen as proposed in Ref.[3]. The cell assembly was then heated at 773-1273 K for 5~7 hours under pressures of 3-5.5 GPa, followed by rapid cooling to room temperature. Finally, $LuH_{2\pm x}N_y$ samples were obtained after pressure release. After the HPHT experiments, we carefully removed the surrounding *h*-BN materials and remaining lutetium on the surfaces (Extended Data Fig. 1b), and then retrieved the final products for characterization and further high-pressure experiments. The obtained HPHT products have a dominated shining blue phase and a very tiny amount of purple phase on the surface. We focus our study on the most interesting blue phase (Extended Data Fig. 1c).

**Sample characterization.** The crystal structure of produced samples was examined using an X-ray diffractometer (XRD) with Cu-Kα radiation ($\lambda$= 1.5406 Å, PANalytical X'pert3, Holland).

High resolution transmission electron microscopy (HRTEM) images were obtained on a JEM2100 Plus transmission electron microscope at an acceleration voltage of 200 KV. Elemental analysis was made by a scanning electron microscope (SEM, Zeiss Sigma 500) equipped with energy dispersive x-ray (EDX) spectroscopy probe using an accelerating voltage of 5 kV. X-ray photoelectron spectroscopy (XPS) spectrums were taken by a Thermo Scientific (ESCALAB, 250Xi). Raman experiments were carried out on a high-resolution Raman spectrometer (Horiba, LabRAM HR revolution) with the excitation wavelength of 532 nm (grating:1800g/mm). The electrical transport measurement was carried out on a Quantum Design Physical Property Measurement System (PPMS).

***In-situ* high pressure measurements.** High pressure resistivity measurements were conducted in a screw-pressure-type diamond-anvil-cell (DAC) made of non-magnetic Be-Cu alloy. A mixture of *c*-BN powder and epoxy was used as the insulating coating for the rhenium gaskets, which was pre-indented to 30 μm in thickness. Several pieces of $LuH_{2\pm x}N_y$ grains with shinning blue color were selected under a microscope and loaded into the gasket hole. The Pt electrodes were attached to the sample with a four-probe van der Pauw method. NaCl was used as the pressure medium. Also, water, silicone oil, and nitrogen gas were used in the experiments of pressure induced color change. High-quality ruby balls ~10 μm in size were used for pressure calibration.

# Supplementary Material

# Observation of non-superconducting phase changes in $LuH_{2\pm x}N_y$

Xiangzhuo Xing[1,2,#], Chao Wang[1,2,#], Linchao Yu[1], Jie Xu[1], Chutong Zhang[1], Mengge Zhang[1], Song Huang[1], Xiaoran Zhang[1], Bingchao Yang[1,2], Xin Chen[1,2], Yongsheng Zhang[1,2], Jian-gang Guo[3], Zhixiang Shi[4], Yanming Ma[5,6,7], Changfeng Chen[8] and Xiaobing Liu[1,2,*]

[1]*Laboratory of High Pressure Physics and Material Science (HPPMS), School of Physics and Physical Engineering, Qufu Normal University, Qufu 273165, China*

[2]*Advanced Research Institute of Multidisciplinary Sciences, Qufu Normal University, Qufu 273165, China*

[3]*Beijing National Laboratory for Condensed Matter Physics, Institute of Physics, Chinese Academy of Sciences, Beijing 100190, China*

[4]*School of Physics, Southeast University, Nanjing 211189, China*

[5]*Innovation Center for Computational Methods & Software, College of Physics, Jilin University, Changchun 130012, China*

[6]*State Key Laboratory of Superhard Materials, Jilin University, Changchun 130012, China*

[7]*International Center of Future Science, Jilin University, Changchun 130012, China*

[8]*Department of Physics and Astronomy, University of Nevada, Las Vegas, Nevada 89154, USA*

[#] *These authors contributed equally to this work.*

[*] Corresponding author: xiaobing.phy@qfnu.edu.cn

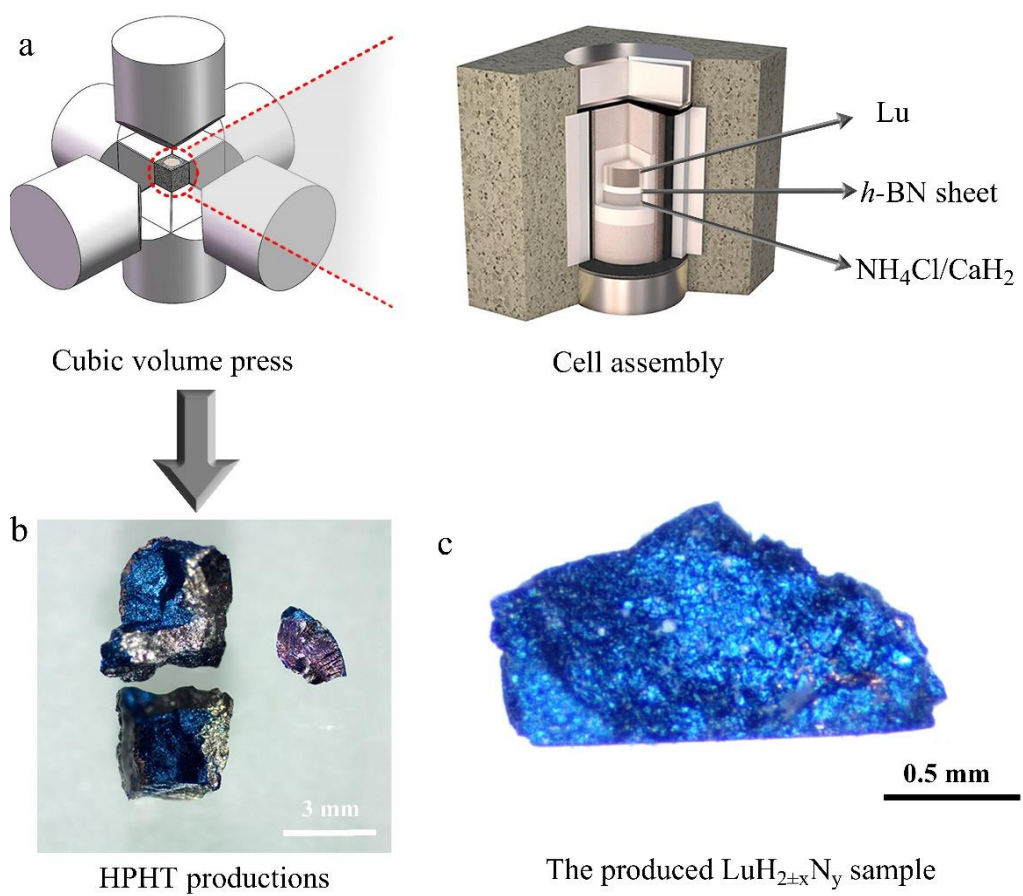

**Extended Data Fig. 1 | Sample synthesis. a**, Schematic illustration of the employed cubic-type high pressure apparatus and HPHT chamber for the synthesis of $LuH_{2\pm x}N_y$ samples. The high-pressure cell has two layers (layer 1 and layer 2) that were separated by a *h*-BN thin plate. In layer 1, the Lu pieces with silver color were placed as precursors. Layer 2 was filled with the mixture of $NH_4Cl$ and $CaH_2$ used as the H/N source. **b**, Optical images of the HPHT products. Most of the as-grown sample surfaces are in silver color. By cracking the sample, the interior shows a uniform shining blue color. **c**, The blue color of the synthesized sample interior is very similar to that of the previously reported $LuH_{3-\delta}N_\epsilon$ at ambient pressure[1].

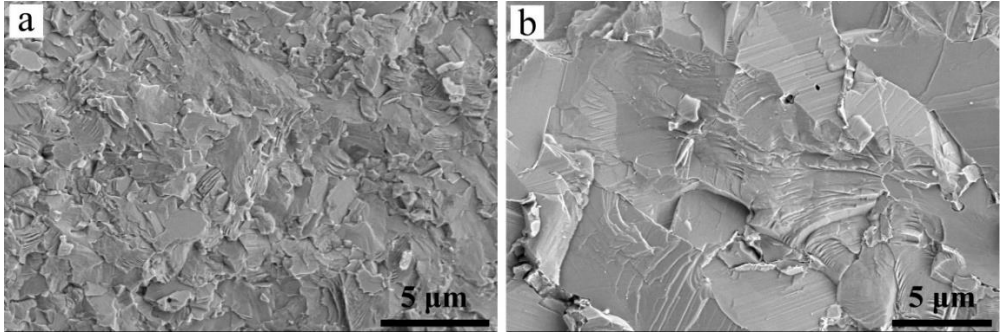

**Extended Data Fig. 2 | Microstructure of the produced LuH$_{2\pm x}$N$_y$ samples. a** and **b**, Typical scanning electron microscope images of the LuH$_{2\pm x}$N$_y$ samples under ambient conditions.

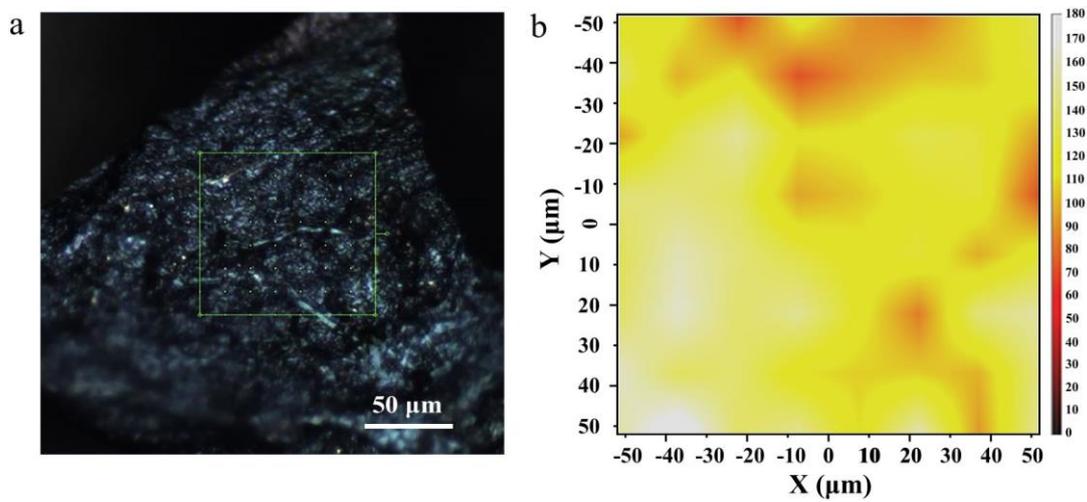

**Extended Data Fig. 3 | Composition characterization for the produced LuH$_{2\pm x}$N$_y$ samples by Raman spectroscopy. a**, Optical image of the area for mapping. **b**, Mapping data of the selected area shows a pure LuH$_{2\pm x}$N$_y$ phase.

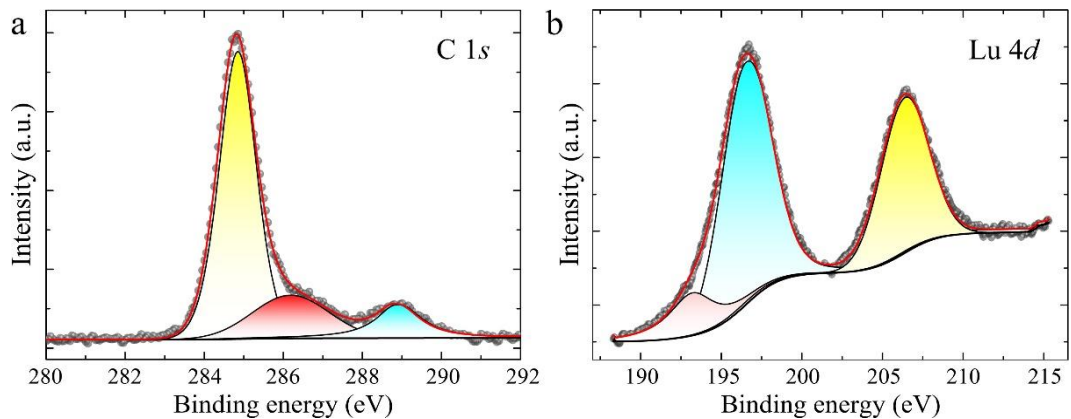

**Extended Data Fig. 4 | Evidence of incorporation of nitrogen in the produced LuH$_{2\pm x}$N$_y$ samples. a** and **b**, X-ray photoelectron spectroscopy (XPS) spectra of C 1$s$ and Lu 4$d$ core levels, respectively. The carbon phase (284.6 eV) from the XPS chamber is used for calibration.

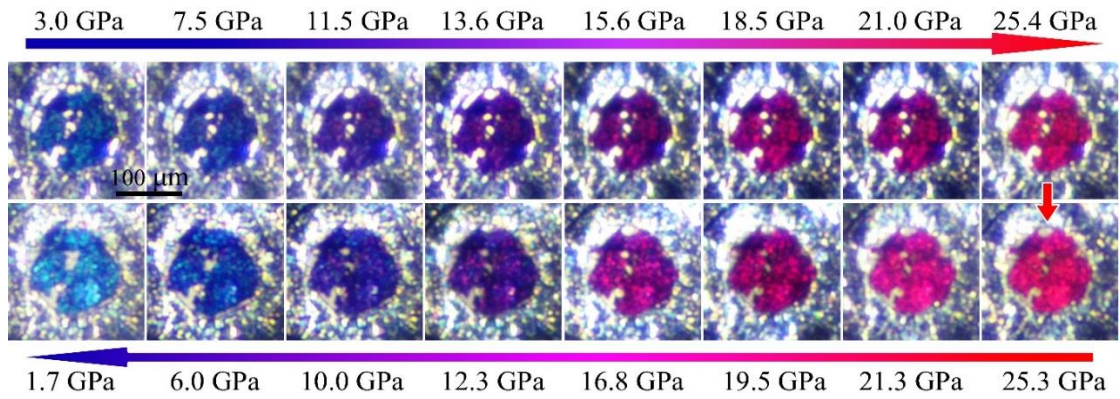

**Extended Data Fig. 5 | Pressure-induced color change of LuH$_{2\pm x}$N$_y$ compressed in nitrogen gas medium.** Optical images of LuH$_{2\pm x}$N$_y$ sample during compression (top row) and decompression (bottom row). As rising pressure, the LuH$_{2\pm x}$N$_y$ sample undergoes a continuous color change from shinning blue at 0 GPa, to purplish blue at 11.5 GPa, to purple at 15.6 GPa, to pink at 18.5 GPa and to bright red at 25.4 GPa. After releasing pressure from 25.4 GPa to ambient pressure, the color gradually recovers to the original shinning blue.

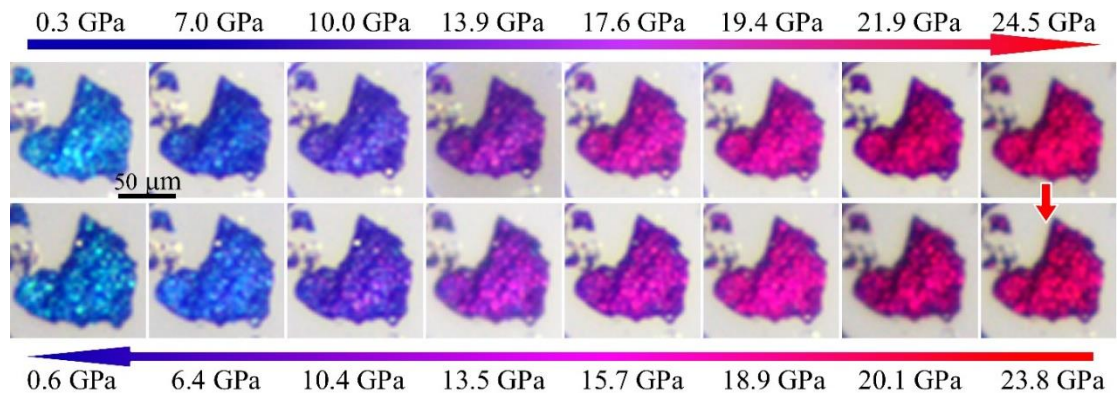

**Extended Data Fig. 6 | Pressure-induced color change of the LuH$_{2\pm x}$N$_y$ sample compressed in water medium.** Optical images of LuH$_{2\pm x}$N$_y$ sample during compression (top row) and decompression (bottom row). At rising pressure, the LuH$_{2\pm x}$N$_y$ sample undergoes a continuous color change from shinning blue at 0 GPa, to purplish blue at 10 GPa, to purple at 13.9 GPa, to pink at 17.6 GPa and to bright red at 21.9 GPa. After releasing pressure from 24.5 GPa to ambient pressure, the color gradually recovers to the original shinning blue.

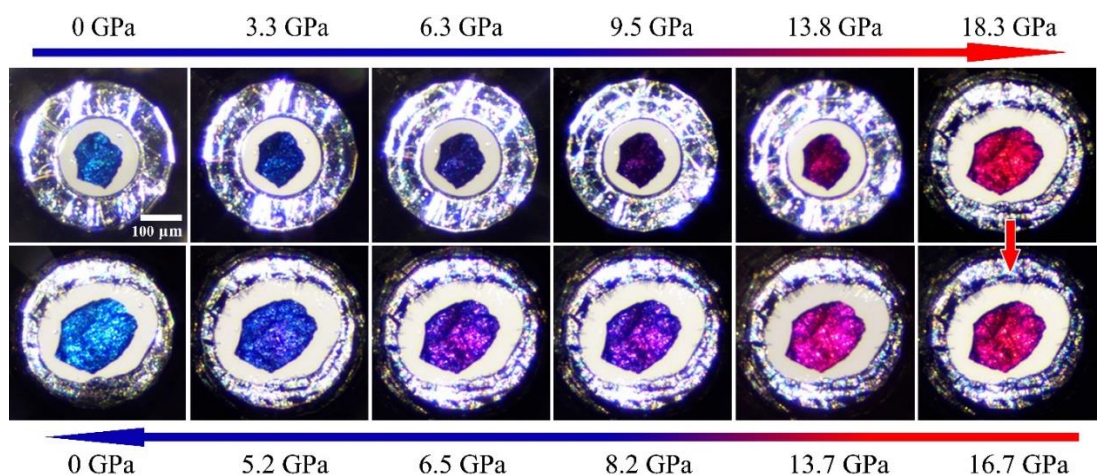

**Extended Data Fig. 7 | Pressure-induced color change of the $LuH_{2\pm x}N_y$ sample compressed in silicone oil medium.** Optical images of $LuH_{2\pm x}N_y$ sample during compression (top row) and decompression (bottom row). At rising pressure, the $LuH_{2\pm x}N_y$ sample undergoes a continuous color change from shinning blue at 0 GPa, to purplish blue at 6.3 GPa, to purple at 9.5 GPa, to pink at 13.8 GPa and to bright red at 18.3 GPa. After releasing pressure from 18.3 GPa to ambient pressure, the color gradually recovers to the original shinning blue.

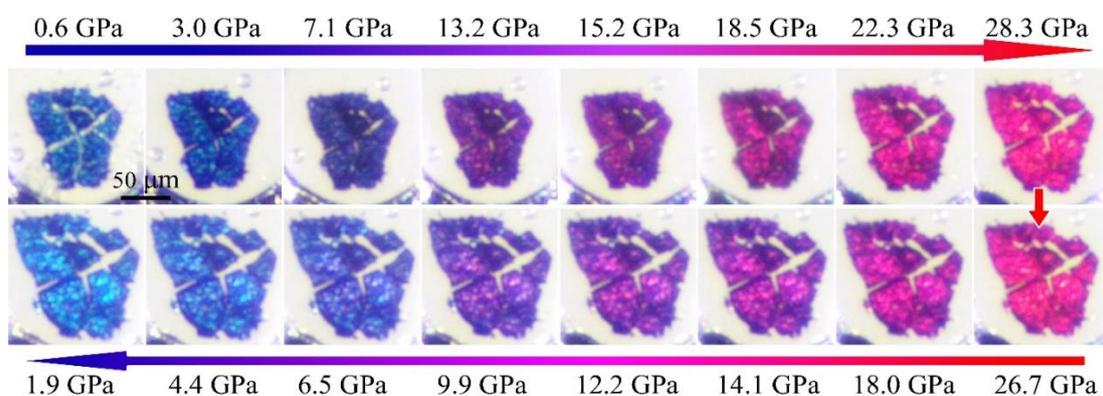

**Extended Data Fig. 8 | Pressure-induced color change of the $LuH_{2\pm x}N_y$ sample compressed in NaCl medium.** Optical images of $LuH_{2\pm x}N_y$ sample during compression (top row) and decompression (bottom row). At rising pressure, the $LuH_{2\pm x}N_y$ sample undergoes a continuous color change from shinning blue at 0 GPa, to purplish blue at 7.1 GPa, to purple at 13.2 GPa, to pink at 18.5 GPa and to bright red (28.3 GPa). After releasing pressure from 28.3 GPa to ambient pressure, the color gradually recovers to the original shinning blue.

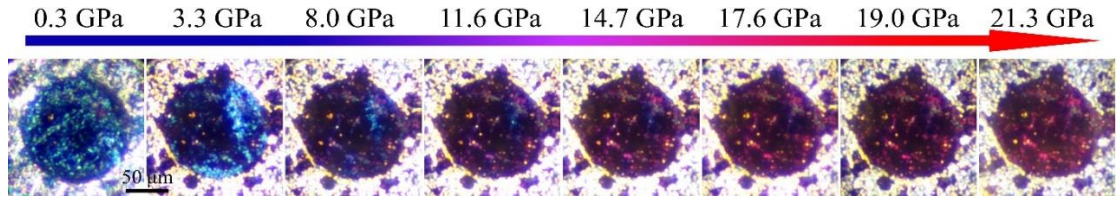

**Extended Data Fig. 9 | Pressure-induced color change of the LuH$_{2\pm x}$N$_y$ sample compressed without any pressure medium.** Optical images of LuH$_{2\pm x}$N$_y$ sample during compression (top row) and decompression (bottom row). A mixed color change with rising pressure is visible because of the inhomogeneous pressure distribution caused by the large pressure gradients in the DAC sample chamber.

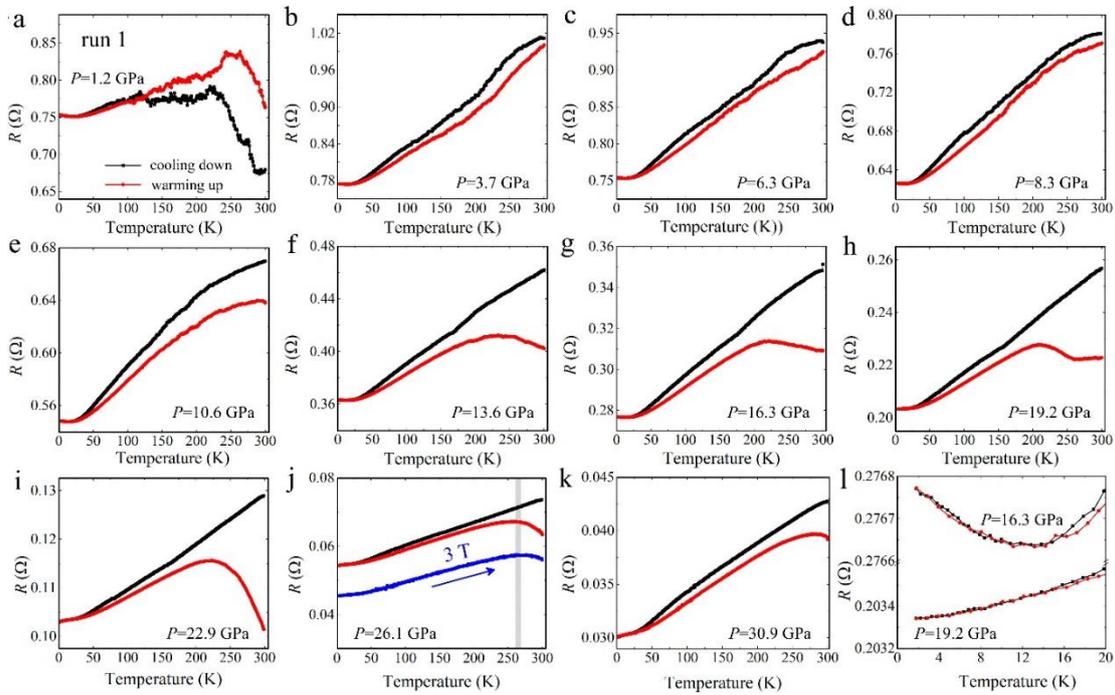

**Extended Data Fig. 10 | Electrical transport measurements under pressure for run 1.** The temperature dependence of resistance of LuH$_{2\pm x}$N$_y$ for run 1 measured in the cooling-down and warming-up processes at different pressures: 1.2 GPa (**a**), 3.7 GPa (**b**), 6.3 GPa (**c**), 8.3 GPa (**d**), 10.6 GPa (**e**), 13.6 GPa (**f**), 16.3 GPa (**g**), 19.2 GPa (**h**), 22.9 GPa (**i**), 26.1 GPa (**j**) and 30.9 GPa (**k**). At $P$=1.2 GPa, the resistance displays a hump feature at high temperatures in both the cooling-down and warming-up processes. As the pressure increases up to 10.6 GPa, the resistance exhibits a monotonous behavior with changing temperature. With further increasing pressure, however, the resistance in the cooling-down and warming-up process exhibit contrasting

behaviors, with an anomalous resistance hump emerging in the warming up process up to 30.9 GPa, which is similar to that observed in a recently reported work[2]. (**l**) A weak resistance upturn is detected in the low temperature region at pressures below 16.3 GPa, which can be suppressed at higher pressures at or above 19.2 GPa. Moreover, to check whether the hump is related to a possible superconducting transition, a magnetic field was applied at $P$ =26.1 GPa, and the results (**j**) show that the application of magnetic field did not move the peak of the resistance curve toward lower temperature as one would expect if the resistance drop was caused by the emergence of superconductivity in the sample.

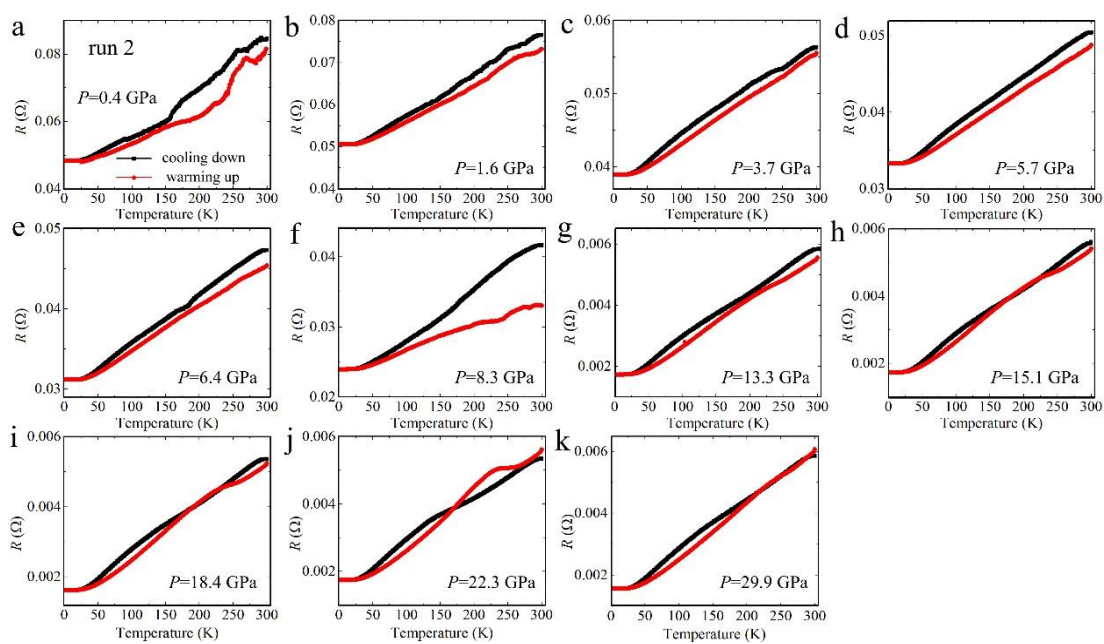

**Extended Data Fig. 11 | Electrical transport measurements under pressure for run 2.** The temperature dependence of resistance of $LuH_{2\pm x}N_y$ for run 2 measured in the cooling-down and warming-up processes at different pressures: 0.4 GPa (**a**), 1.6 GPa (**b**), 3.7 GPa (**c**), 5.7 GPa (**d**), 6.4 GPa (**e**), 8.3 GPa (**f**), 13.3 GPa (**g**), 15.1 GPa (**h**), 18.4 GPa (**i**), 22.3 GPa (**j**) and 29.9 GPa (**k**). At low pressures, the resistance shows similar behaviors with those seen in run 1. At $P$=8.3 GPa, the resistance in the cooling-down and warming-up processes starts to show different behaviors, followed by a large (an order of magnitude) drop at around 13.3 GPa, as described in the main text. At higher pressures above 13.3 GPa, a resistance kink at high temperatures is observed in the warming-up process, which probably share the same origin that drives the phase change indicated by the hump seen in run 1, despite with the different curve shapes. At low temperatures, a weak

upturn is also observed below 5.7 GPa, but is absent at higher pressures.

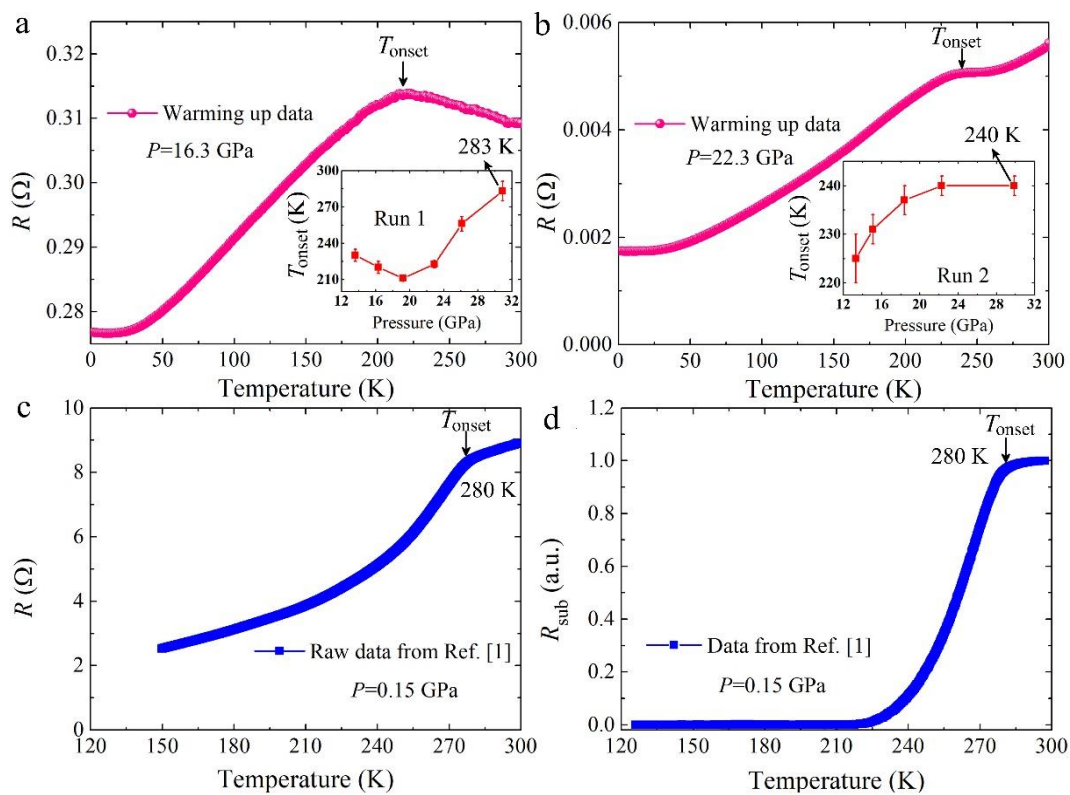

**Extended Data Fig. 12 | Comparison of the resistance anomaly in $LuH_{2\pm x}N_y$ and the claimed superconducting transition[1].** As described in the main text, a resistance anomaly with a hump/plateau at high temperatures has been observed during the warming up process, with typical results shown in **a** and **b**. Such anomaly starts to appear in the purple phase and persists through the pink and red phases, possibly due to a structural/electronic phase transition. As the temperature increases, the characteristic temperature $T_{onset}$ labelled by the black arrows exhibits a rising trend in general through the pink phase, as shown in the insets of **a** and **b**. It is seen that the warming-driven behavior of resistance anomaly below $T_{onset}$ is very similar to that reported for near-ambient superconductors[1], especially for the pink phase at 22.3 GPa in our run 2. For comparison, the source data were downloaded from Ref. [1] and plotted in **c**, from which the superconducting state with zero resistance was achieved after subtracting a so-called background signal, as shown in **d**.